\documentclass[lengthestimate,tighten,amssymb,prl,aps,twocolumn,floatfix,tightenlines,superscriptaddress]{revtex4}
\usepackage{epsfig}
\usepackage{amsfonts}
\usepackage{amsmath}
\usepackage{epsfig}

\begin{document}

\title{An All Optical Fibre Quantum Controlled-NOT Gate}
\author{Alex S. Clark}
\affiliation{Centre for Quantum Photonics, H. H. Wills Physics Laboratory \& Department of Electrical and Electronic Engineering, University of Bristol, Merchant Venturers Building, Woodland Road, Bristol, BS8 1UB, UK}
\author{J\'er\'emie Fulconis}
\affiliation{Centre for Quantum Photonics, H. H. Wills Physics Laboratory \& Department of Electrical and Electronic Engineering, University of Bristol, Merchant Venturers Building, Woodland Road, Bristol, BS8 1UB, UK}
\author{John G. Rarity}
\affiliation{Centre for Quantum Photonics, H. H. Wills Physics Laboratory \& Department of Electrical and Electronic Engineering, University of Bristol, Merchant Venturers Building, Woodland Road, Bristol, BS8 1UB, UK}
\author{William J. Wadsworth}
\affiliation{Centre for Photonics and Photonic Materials, Department of Physics, University of Bath, Claverton Down, Bath, BA2 7AY, UK}
\author{Jeremy L. O'Brien}
\affiliation{Centre for Quantum Photonics, H. H. Wills Physics Laboratory \& Department of Electrical and Electronic Engineering, University of Bristol, Merchant Venturers Building, Woodland Road, Bristol, BS8 1UB, UK}

\begin{abstract}%
We report the first experimental demonstration of an optical controlled-NOT gate constructed entirely in fibre. We operate the gate using two heralded optical fibre single photon sources and find an average logical fidelity of $90\%$ and an average process fidelity of $0.83\leq \overline{F}\leq 0.91$. On the basis of a simple model we are able to conclude that imperfections are primarily due to the photon sources, meaning that the gate itself works with very high fidelity.
\end{abstract}
\date{\today}
\maketitle

Quantum information science aims to harness uniquely quantum mechanical effects to greatly improve performance and functionality in the encoding, transmission and processing of information. Single photons are the logical choice for quantum communication \cite{gi-rmp-74-145}, quantum metrology \cite{na-sci-316-726,hi-nat} and lithography \cite{da-prl-87-013602}, and are a leading approach to quantum information processing \cite{ob-sci-318-1567}. There have been a number of impressive proof-of-principle demonstrations of photonic quantum circuits in each of these areas, however, to date they have relied on large-scale optical elements with photons propagating in air, making future practical applications difficult. Here we demonstrate an all fibre implementation of a controlled-NOT (CNOT) gate using two heralded single photon sources.  We measure an average logical fidelity of $90\%$ and an average process fidelity of $0.83\leq \overline{F}\leq 0.91$. Using a simple model we find the remaining discrepancy to be due almost entirely to spectral properties of the heralded sources, demonstrating near-perfect operation of the fibre CNOT gate itself.

\begin{figure}[b!]
\begin{center}
%\vspace{-0.3cm}
\includegraphics*[width=0.45\textwidth]{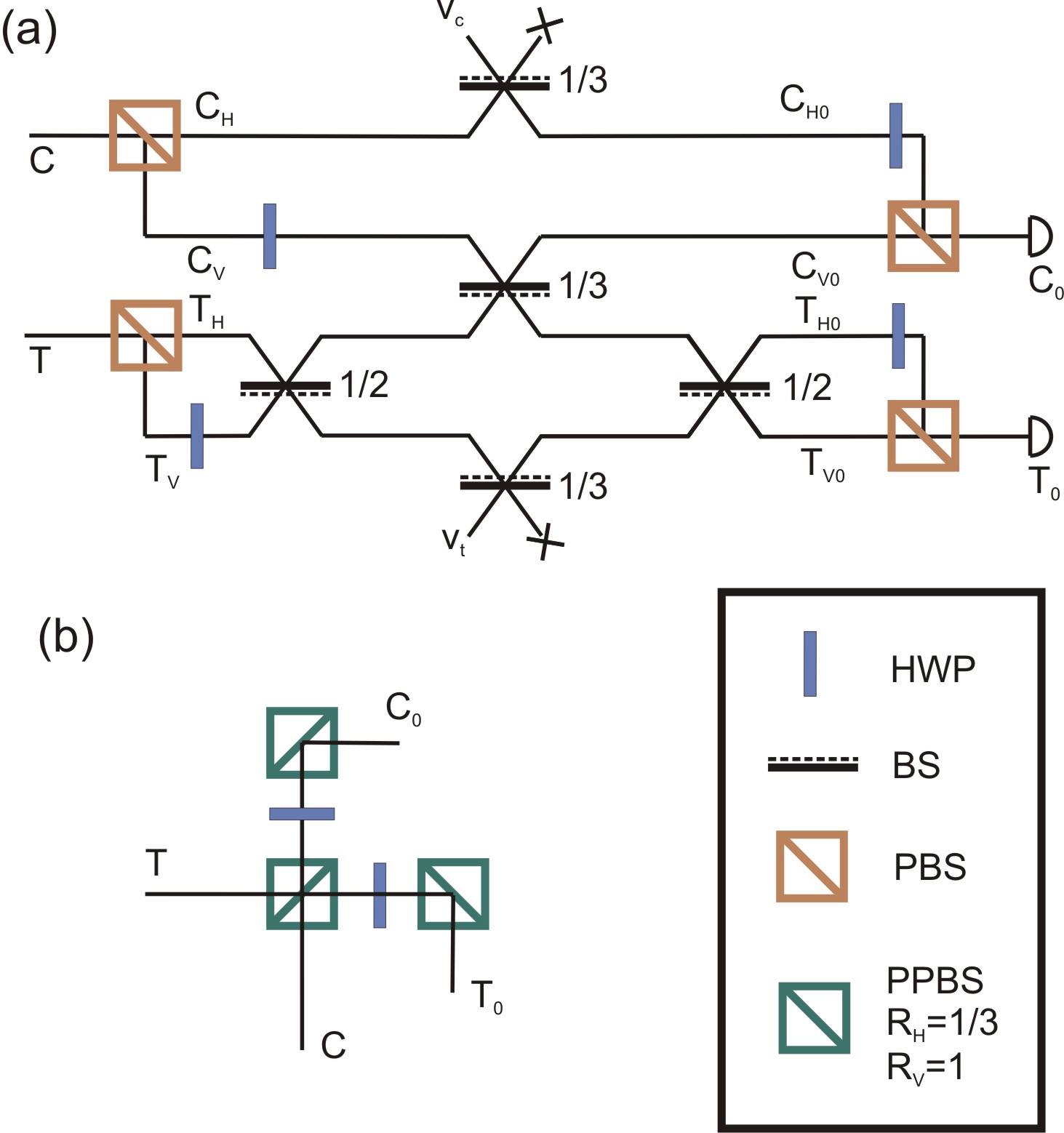}
\vspace{-0cm}
\caption{A two-photon quantum CNOT gate. (a) The setup involves conversion from polarization encoding to spatial mode encoding via the use of polarizing beamsplitters (PBS) in the control, C, and target, T, modes followed by a non-classical interference on the central beamsplitter with a reflectivity equal to 1/3; two additional beamsplitters ensure correct output amplitudes, with a dotted line indicating a sign change on reflection.  The beams are then recombined at further polarizing beamsplitters, creating multiple interferometers.  Success of the gate is conditional upon detecting a photon at each of the target and control outputs, which occurs with probability 1/9. (b) The PPBS simplification of the gate.
}
\label{original}
\end{center}
\end{figure}

A major breakthrough showed that the interactions between photons required for photonic quantum technologies could be realized using a measurement induced optical non-linearity \cite{kn-nat-409-46}. 
An important circuit built on such an interaction is the CNOT gate shown in Fig. 1(a), which flips the state of the target qubit conditional on the control qubit being in the logical ``1" state \cite{ra-pra-65-062324,ho-pra-66-024308}. The original demonstration of this circuit relied on a sophisticated interferometer to maintain the stable phase required in converting between polarization and path encoding \cite{ob-nat-426-264,ob-prl-93-080502}. This complication was circumvented by schemes \cite{la-prl-95-210504,ki-prl-95-210505,ok-prl-95-210506} which used ``partially polarizing beamsplitters" (PPBS) with different reflectivities for each polarization: 1 and 1/3 for vertically and horizontally polarized photons respectively [Fig. 1(b)]. However, these demonstrations have all relied on large scale (bulk) optical elements bolted to optical tables, with photons propagating in air. Practical applications of these and other photonic quantum circuits will likely require a guided optical implementation to enable improved performance, miniaturization, and scalability.

We use the guided optical network of three partially polarizing fibre couplers (PPFCs) shown schematically in Fig. 2, to implement a two photon CNOT gate. 
 \begin{figure*}[t!]
\begin{center}
%\vspace{0.3cm}
\includegraphics[width=0.9\textwidth]{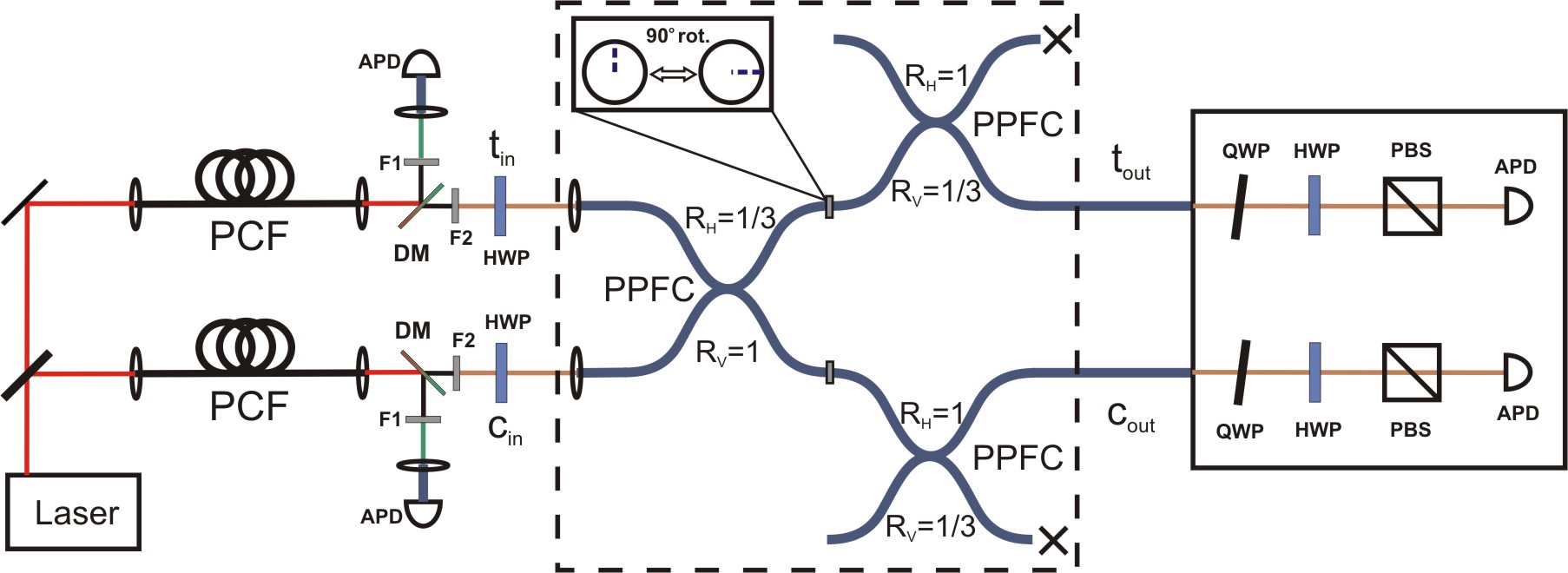}
%\vspace{0.3cm}
\caption{An all-optical fibre quantum CNOT gate with heralded single photon sources. A ps 708 nm Ti:sapphire laser pumps two photonic crystal fibres (PCFs) creating a non-degenerate pair of photons at 583 nm and 900 nm.  These are seperated at dichroic mirrors (DM), and pass through interference filters F1 and F2 with bandwidths of 0.2 nm and 0.8 nm respectively.  Detection of the 583 nm photons heralds the arrival of the 900 nm control C and target T photons at the inputs to the CNOT gate. Half-wave plates (HWPs) are used to create logical and diagonal input states.  The gate consists of three partially polarizing fibre couplers (PPFC) with the specific reflectivities for horizontal and vertical photons as shown.  The polarizations of the idler photons are then analysed using a HWP and a polarizing beam splitter (PBS) cube for each arm.  Note that the Hadamard operations before and after the gate are integrated into the encoding and analysis wave plates.  Quarter-wave plates (QWPs) set to 0$^{\circ}$ are tilted to correct for the phase accumulated in the polarization maintaining fibres of the gate.  All four photons are detected using Perkin Elmer silicon avalanche photodiodes (APD's) and sent to electronic 4-fold coincidence counting circuitry for analysis.}
%\vspace{0.3cm}
\label{schematic}
\end{center}
\end{figure*} 
As with the previous bulk optical PPBS implementations, this circuit retains the polarization encoding throughout, removing the need for classical interference. In addition it guarantees excellent spatial mode-matching required to realize high-fidelity quantum interference, allows the gate to be miniaturized, and exhibits the criteria for scalable and integrable photonic quantum circuits.

A schematic of the experimental setup is shown in Fig. \ref{schematic}. 
We used two heralded photon sources, consisting of two photonic crystal fibres \cite{fu-prl-99-120501}, to signal the arrival of the control and target photons. The CNOT gate is constructed from three partially polarizing fibre couplers (PPFCs): polarization maintaining fibre couplers with reflectivities $R_H=1/3$ and $R_V=1$. The first PPFC enables the control c and target t photons to interact via quantum interference and subsequent photon detection. The other two PPFCs serve to balance the quantum amplitudes and thereby maintain a uniform success probability of 1/9 \cite{ra-pra-65-062324,ho-pra-66-024308}. Polarization maintaining fibres allow two orthogonal polarizations (those parallel to the axes of birefringence) to propagate without rotation (horizontal and vertical in our case). These two polarizations travel at a different speed through the fibre, such that a superposition state will be rotated and ultimately decohere.  This is coarsely corrected for within the gate by joining the PPFCs with a 90$^{\circ}$ rotation at the connections (Fig. 2). This rotation also has the effect of swapping the reflection coefficients  $R_{H}\leftrightarrow R_{V}$, as required. The remaining phase shift is corrected by a tilted birefringent waveplate at each output of the gate. Multi-photon contributions are subtracted from the data \cite{fu-prl-99-120501}.

\begin{figure*}[t!]
\begin{center}
\vspace{0.5cm}
\includegraphics*[width=1\textwidth]{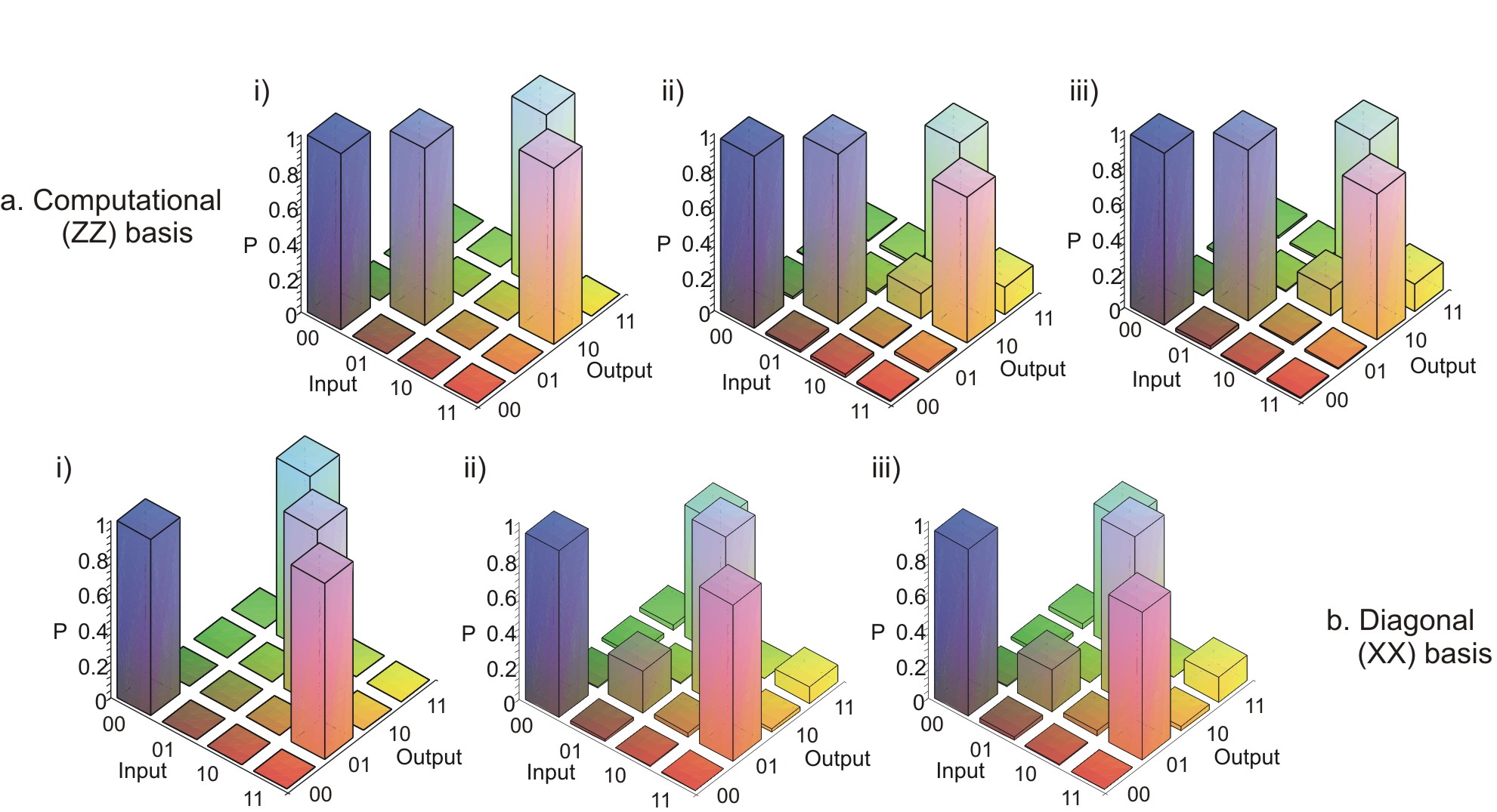}
\vspace{0.3cm}
\caption{Truth table results for the gate operating in two orthogonal bases, (a) the computational (ZZ) basis and (b) the diagonal (XX) basis.  We show (i) the ideal operation, (ii) the experimental data and  (iii) the results from the model.}
\vspace{0.3cm}
\label{results}
\end{center}
\end{figure*}

Following the method proposed in Ref. \onlinecite{ho-prl-94-160504} and implemented in Ref{. }\onlinecite{ok-prl-95-210506}, we measure truth tables for two complementary bases.  We choose the computational (ZZ) basis: $\{|0\rangle=|V\rangle,|1\rangle=|H\rangle\}$; and the diagonal (XX) basis: $\{|0\rangle=(|H\rangle+|V\rangle)/\sqrt{2},|1\rangle=(|H\rangle-|V\rangle)/\sqrt{2}\}$. The 16 element truth tables are  defined as the probability of obtaining each of the four logical output states for each logical input. The results are shown in Fig. \ref{results}.  From these data we calculate average logical fidelities $F_{ZZ}$=0.90$\pm$0.02 for the computational (ZZ) basis, and $F_{XX}$=0.89$\pm$0.02 for the diagonal (XX) basis. These two classical fidelities can be used to place bounds on the quantum process fidelity $F_{P}$ \cite{ho-prl-94-160504}:  
\begin{equation}
\label{Bound}
F_{XX}+F_{ZZ}-1\leq F_{P}\leq Min\{F_{XX},F_{ZZ}\}
\end{equation}
\noindent which gives $0.79\leq F_{P}\leq0.89$ for our all-fibre gate. This corresponds to an average process fidelity \cite{ob-prl-93-080502,gi-pra-71-062310} of $0.83\leq \overline{F}\leq 0.91$

\begin{figure}[b!]
\begin{center}
\includegraphics*[width=0.45\textwidth]{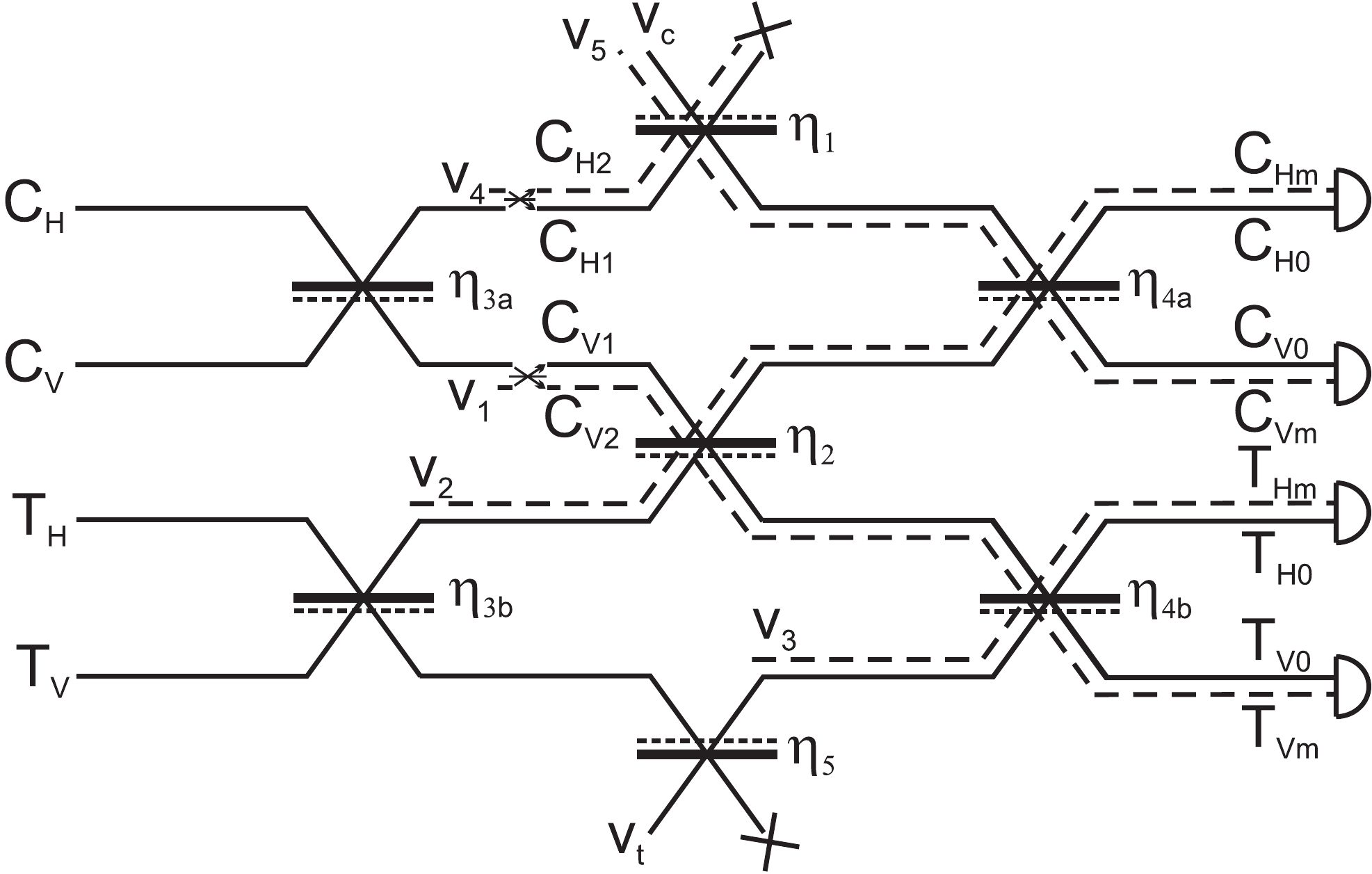}
\caption{A model for CNOT gate operation. The additional modes (dashed lines) are used to model mode mismatch at the central beamsplitter. The beamsplitters are given reflectivites $\eta_{i}$.  The dotted-line side of the beamsplitters indicates the side at which a $\pi$ phase shift occurs on reflection.}
\label{model}
\end{center}
\end{figure}

The quantum interference visibility $V$ at the central PPFC is 94\%, most likely due to spectral mismatch of the photons \cite{fu-prl-99-120501}. To understand and quantify the effect of this error we have developed a model based on Ref. \onlinecite{ra-pra-65-062324} which also includes imperfect state preparation and analysis (Fig. \ref{model}). The mode mismatch is modelled by introduction of two extra spatial modes $C_{H2}$, $C_{V2}$; although this is modelled as spatial, mismatch in all degrees of freedom are equivalent \cite{ro-pra-72-032306}. The beamsplitter values in the model are determined directly from measurement of the PPFC reflectivities. Incorrect state preparation and analysis is modelled using beamsplitters which mix the target and control H and V modes.  Ideal operation in the computational basis corresponds to $\eta_{3a}=\eta_{4a}=1$ and $\eta_{3b}=\eta_{4b}=1/2$; and in the diagonal basis $\eta_{3a}=\eta_{4a}=1/2$ and $\eta_{3b}=\eta_{4b}=1$ (\emph{i.e.} control and target are swapped).  %By varying these reflectivities we see the elements of the truth table take on non-ideal values.  
Imperfect quantum interference results in (equal) error terms on the diagonal of the truth table for the control ``1" inputs, in both bases. Truth tables from the full model are shown in Fig. \ref{results}.

As it is not possible to use the average fidelity to measure the distance between two non-ideal truth tables, we introduce the similarity $S$ to compare the truth table generated by the model $M$ and that measured experimentally $E$:
\begin{equation}
\label{compare}
S=\frac{\left(\displaystyle\sum_{i,j=1}^{4}\sqrt{M_{i,j}E_{i,j}}\right)^2}{16}
\end{equation}
\noindent
which is a generalisation of the average fidelity based on the (classical) fidelity between probability distributions \cite{pr-prl-92-190402,ra-pra-73-012113}.

The similarity of the experimental data to the model is shown in Table \ref{modeltable} for three different cases: ideal operation; the case where only imperfect quantum interference is included; and the case where imperfect encoding and decoding are also incorporated. %shall we take em out?
The small deviations of the beamsplitter values from the ideal case proved to have a negligible effect on the fidelity, but were retained for completeness.  The results show that while there were some small errors introduced through the waveplate positioning, most of the errors, $7.9\%$ and $7.7\%$ for the computational and diagonal bases respectively, were due to an imperfect interference visibility.  The source of mode mismatch is unlikely to be spatial, as it has been shown previously that fibre beamsplitters exhibit interference visibilities of greater than $99\%$ \cite{pi-prl-90-240401}. It is more likely that the mismatch arises through spectral differences between the photons emitted from the seperate sources \cite{fu-prl-99-120501}. Introducing imperfect state preparation and analysis causes the similarity of the model with the experimental results to rise by a further $2.3\%$ and $3.8\%$, reaching $99.7\%$ and $99.8\%$ for the computational and diagonal bases respectively, confirming near-perfect operation of the CNOT gate itself.

\begin{table}
\begin{tabular}{| c | c  c  c | c  c  c | }
\hline			
  MODEL TYPE & & $S_{ZZ}$ & & & $S_{XX}$ & \\
  \hline
  IDEAL & & 0.895 & & & 0.883 & \\
  INTERFERENCE & & 0.974 & & & 0.960 & \\
  FULL MODEL & & 0.997 & & & 0.998 & \\
\hline
\end{tabular}
\caption{Similarity between the model and experimental results for the ideal gate, the case with imperfect visibility of the quantum interference and the full model including imperfect visibility and imperfect state preparation and analysis. \label{modeltable}}
\end{table}

We have demonstrated a high-fidelity guided optical implementation of a two-photon CNOT gate operating on polarization. This opens the way for the development of integrated quantum optical devices and the miniaturization of photonic quantum circuits for future technologies as well as fundamental studies in quantum optics.\\

W.J.W. is a Royal Society University Research Fellow, J.G.R. is supported by a Wolfson Merit award. The work was supported by the UK EPSRC (QIP IRC and 1-phot), the US Disruptive Technologies Office, EU IP QAP and FP6-2002-IST-1-506813 SECOQC.


\begin{thebibliography}{20}
\expandafter\ifx\csname natexlab\endcsname\relax\def\natexlab#1{#1}\fi
\expandafter\ifx\csname bibnamefont\endcsname\relax
  \def\bibnamefont#1{#1}\fi
\expandafter\ifx\csname bibfnamefont\endcsname\relax
  \def\bibfnamefont#1{#1}\fi
\expandafter\ifx\csname citenamefont\endcsname\relax
  \def\citenamefont#1{#1}\fi
\expandafter\ifx\csname url\endcsname\relax
  \def\url#1{\texttt{#1}}\fi
\expandafter\ifx\csname urlprefix\endcsname\relax\def\urlprefix{URL }\fi
\providecommand{\bibinfo}[2]{#2}
\providecommand{\eprint}[2][]{\url{#2}}

\bibitem[{\citenamefont{Gisin et~al.}(2002)\citenamefont{Gisin, Ribordy,
  Tittel, and Zbinden}}]{gi-rmp-74-145}
\bibinfo{author}{\bibfnamefont{N.}~\bibnamefont{Gisin}},
  \bibinfo{author}{\bibfnamefont{G.}~\bibnamefont{Ribordy}},
  \bibinfo{author}{\bibfnamefont{W.}~\bibnamefont{Tittel}}, \bibnamefont{and}
  \bibinfo{author}{\bibfnamefont{H.}~\bibnamefont{Zbinden}},
  \bibinfo{journal}{Rev. Mod. Phys.} \textbf{\bibinfo{volume}{74}},
  \bibinfo{pages}{145} (\bibinfo{year}{2002}).

\bibitem[{\citenamefont{Nagata et~al.}(2007)\citenamefont{Nagata, Okamoto,
  O'Brien, Sasaki, and Takeuchi}}]{na-sci-316-726}
\bibinfo{author}{\bibfnamefont{T.}~\bibnamefont{Nagata}},
  \bibinfo{author}{\bibfnamefont{R.}~\bibnamefont{Okamoto}},
  \bibinfo{author}{\bibfnamefont{J.~L.} \bibnamefont{O'Brien}},
  \bibinfo{author}{\bibfnamefont{K.}~\bibnamefont{Sasaki}}, \bibnamefont{and}
  \bibinfo{author}{\bibfnamefont{S.}~\bibnamefont{Takeuchi}},
  \bibinfo{journal}{Science} \textbf{\bibinfo{volume}{316}},
  \bibinfo{pages}{726} (\bibinfo{year}{2007}).

\bibitem[{\citenamefont{Higgins et~al.}(2007)\citenamefont{Higgins, Berry,
  Bartlett, Wiseman, and Pryde}}]{hi-nat}
\bibinfo{author}{\bibfnamefont{B.~L.} \bibnamefont{Higgins}},
  \bibinfo{author}{\bibfnamefont{D.~W.} \bibnamefont{Berry}},
  \bibinfo{author}{\bibfnamefont{S.~D.} \bibnamefont{Bartlett}},
  \bibinfo{author}{\bibfnamefont{H.~M.} \bibnamefont{Wiseman}},
  \bibnamefont{and} \bibinfo{author}{\bibfnamefont{G.~J.} \bibnamefont{Pryde}},
  \bibinfo{journal}{Nature} \textbf{\bibinfo{volume}{450}},
  \bibinfo{pages}{393} (\bibinfo{year}{2007}).

\bibitem[{\citenamefont{D'Angelo et~al.}(2001)\citenamefont{D'Angelo, Chekhova,
  and Shih}}]{da-prl-87-013602}
\bibinfo{author}{\bibfnamefont{M.}~\bibnamefont{D'Angelo}},
  \bibinfo{author}{\bibfnamefont{M.~V.} \bibnamefont{Chekhova}},
  \bibnamefont{and} \bibinfo{author}{\bibfnamefont{Y.}~\bibnamefont{Shih}},
  \bibinfo{journal}{Phys. Rev. Lett.} \textbf{\bibinfo{volume}{87}},
  \bibinfo{pages}{013602} (\bibinfo{year}{2001}).

\bibitem[{\citenamefont{O'Brien}(2007)}]{ob-sci-318-1567}
\bibinfo{author}{\bibfnamefont{J.~L.} \bibnamefont{O'Brien}},
  \bibinfo{journal}{Science} \textbf{\bibinfo{volume}{318}},
  \bibinfo{pages}{1567} (\bibinfo{year}{2007}).

\bibitem[{\citenamefont{Knill et~al.}(2001)\citenamefont{Knill, Laflamme, and
  Milburn}}]{kn-nat-409-46}
\bibinfo{author}{\bibfnamefont{E.}~\bibnamefont{Knill}},
  \bibinfo{author}{\bibfnamefont{R.}~\bibnamefont{Laflamme}}, \bibnamefont{and}
  \bibinfo{author}{\bibfnamefont{G.~J.} \bibnamefont{Milburn}},
  \bibinfo{journal}{Nature} \textbf{\bibinfo{volume}{409}}, \bibinfo{pages}{46}
  (\bibinfo{year}{2001}).

\bibitem[{\citenamefont{Ralph et~al.}(2001)\citenamefont{Ralph, Langford, Bell,
  and White}}]{ra-pra-65-062324}
\bibinfo{author}{\bibfnamefont{T.~C.} \bibnamefont{Ralph}},
  \bibinfo{author}{\bibfnamefont{N.~K.} \bibnamefont{Langford}},
  \bibinfo{author}{\bibfnamefont{T.~B.} \bibnamefont{Bell}}, \bibnamefont{and}
  \bibinfo{author}{\bibfnamefont{A.~G.} \bibnamefont{White}},
  \bibinfo{journal}{Phys. Rev. A} \textbf{\bibinfo{volume}{65}},
  \bibinfo{pages}{062324} (\bibinfo{year}{2001}).

\bibitem[{\citenamefont{Hofmann and Takeuchi}(2001)}]{ho-pra-66-024308}
\bibinfo{author}{\bibfnamefont{H.~F.} \bibnamefont{Hofmann}} \bibnamefont{and}
  \bibinfo{author}{\bibfnamefont{S.}~\bibnamefont{Takeuchi}},
  \bibinfo{journal}{Phys. Rev. A} \textbf{\bibinfo{volume}{66}},
  \bibinfo{pages}{024308} (\bibinfo{year}{2001}).

\bibitem[{\citenamefont{O'Brien et~al.}(2003)\citenamefont{O'Brien, Pryde,
  White, Ralph, and Branning}}]{ob-nat-426-264}
\bibinfo{author}{\bibfnamefont{J.~L.} \bibnamefont{O'Brien}},
  \bibinfo{author}{\bibfnamefont{G.~J.} \bibnamefont{Pryde}},
  \bibinfo{author}{\bibfnamefont{A.~G.} \bibnamefont{White}},
  \bibinfo{author}{\bibfnamefont{T.~C.} \bibnamefont{Ralph}}, \bibnamefont{and}
  \bibinfo{author}{\bibfnamefont{D.}~\bibnamefont{Branning}},
  \bibinfo{journal}{Nature} \textbf{\bibinfo{volume}{426}},
  \bibinfo{pages}{264} (\bibinfo{year}{2003}).

\bibitem[{\citenamefont{O'Brien et~al.}(2004)\citenamefont{O'Brien, Pryde,
  Gilchrist, James, Langford, Ralph, and White}}]{ob-prl-93-080502}
\bibinfo{author}{\bibfnamefont{J.~L.} \bibnamefont{O'Brien}},
  \bibinfo{author}{\bibfnamefont{G.~J.} \bibnamefont{Pryde}},
  \bibinfo{author}{\bibfnamefont{A.}~\bibnamefont{Gilchrist}},
  \bibinfo{author}{\bibfnamefont{D.~F.~V.} \bibnamefont{James}},
  \bibinfo{author}{\bibfnamefont{N.~K.} \bibnamefont{Langford}},
  \bibinfo{author}{\bibfnamefont{T.~C.} \bibnamefont{Ralph}}, \bibnamefont{and}
  \bibinfo{author}{\bibfnamefont{A.~G.} \bibnamefont{White}},
  \bibinfo{journal}{Phys. Rev. Lett.} \textbf{\bibinfo{volume}{93}},
  \bibinfo{eid}{080502} (\bibinfo{year}{2004}).

\bibitem[{\citenamefont{Langford et~al.}(2005)\citenamefont{Langford, Weinhold,
  Prevedel, Resch, Gilchrist, O'Brien, Pryde, and White}}]{la-prl-95-210504}
\bibinfo{author}{\bibfnamefont{N.~K.} \bibnamefont{Langford}},
  \bibinfo{author}{\bibfnamefont{T.~J.} \bibnamefont{Weinhold}},
  \bibinfo{author}{\bibfnamefont{R.}~\bibnamefont{Prevedel}},
  \bibinfo{author}{\bibfnamefont{K.~J.} \bibnamefont{Resch}},
  \bibinfo{author}{\bibfnamefont{A.}~\bibnamefont{Gilchrist}},
  \bibinfo{author}{\bibfnamefont{J.~L.} \bibnamefont{O'Brien}},
  \bibinfo{author}{\bibfnamefont{G.~J.} \bibnamefont{Pryde}}, \bibnamefont{and}
  \bibinfo{author}{\bibfnamefont{A.~G.} \bibnamefont{White}},
  \bibinfo{journal}{Phys. Rev. Lett.} \textbf{\bibinfo{volume}{95}},
  \bibinfo{eid}{210504} (\bibinfo{year}{2005}).

\bibitem[{\citenamefont{Kiesel et~al.}(2005)\citenamefont{Kiesel, Schmid,
  Weber, Ursin, and Weinfurter}}]{ki-prl-95-210505}
\bibinfo{author}{\bibfnamefont{N.}~\bibnamefont{Kiesel}},
  \bibinfo{author}{\bibfnamefont{C.}~\bibnamefont{Schmid}},
  \bibinfo{author}{\bibfnamefont{U.}~\bibnamefont{Weber}},
  \bibinfo{author}{\bibfnamefont{R.}~\bibnamefont{Ursin}}, \bibnamefont{and}
  \bibinfo{author}{\bibfnamefont{H.}~\bibnamefont{Weinfurter}},
  \bibinfo{journal}{Phys. Rev. Lett.} \textbf{\bibinfo{volume}{95}},
  \bibinfo{eid}{210505} (\bibinfo{year}{2005}).

\bibitem[{\citenamefont{Okamoto et~al.}(2005)\citenamefont{Okamoto, Hofmann,
  Takeuchi, and Sasaki}}]{ok-prl-95-210506}
\bibinfo{author}{\bibfnamefont{R.}~\bibnamefont{Okamoto}},
  \bibinfo{author}{\bibfnamefont{H.~F.} \bibnamefont{Hofmann}},
  \bibinfo{author}{\bibfnamefont{S.}~\bibnamefont{Takeuchi}}, \bibnamefont{and}
  \bibinfo{author}{\bibfnamefont{K.}~\bibnamefont{Sasaki}},
  \bibinfo{journal}{Phys. Rev. Lett.} \textbf{\bibinfo{volume}{95}},
  \bibinfo{eid}{210506} (\bibinfo{year}{2005}).

\bibitem[{\citenamefont{Fulconis et~al.}(2007)\citenamefont{Fulconis, Alibart,
  O'Brien, Wadsworth, and Rarity}}]{fu-prl-99-120501}
\bibinfo{author}{\bibfnamefont{J.}~\bibnamefont{Fulconis}},
  \bibinfo{author}{\bibfnamefont{O.}~\bibnamefont{Alibart}},
  \bibinfo{author}{\bibfnamefont{J.~L.} \bibnamefont{O'Brien}},
  \bibinfo{author}{\bibfnamefont{W.~J.} \bibnamefont{Wadsworth}},
  \bibnamefont{and} \bibinfo{author}{\bibfnamefont{J.~G.}
  \bibnamefont{Rarity}}, \bibinfo{journal}{Physical Review Letters}
  \textbf{\bibinfo{volume}{99}}, \bibinfo{eid}{120501} (\bibinfo{year}{2007}).

\bibitem[{\citenamefont{Hofmann}(2005)}]{ho-prl-94-160504}
\bibinfo{author}{\bibfnamefont{H.~F.} \bibnamefont{Hofmann}},
  \bibinfo{journal}{Phys. Rev. Lett.} \textbf{\bibinfo{volume}{94}},
  \bibinfo{eid}{160504} (\bibinfo{year}{2005}).

\bibitem[{\citenamefont{Gilchrist et~al.}(2005)\citenamefont{Gilchrist,
  Langford, and Nielsen}}]{gi-pra-71-062310}
\bibinfo{author}{\bibfnamefont{A.}~\bibnamefont{Gilchrist}},
  \bibinfo{author}{\bibfnamefont{N.~K.} \bibnamefont{Langford}},
  \bibnamefont{and} \bibinfo{author}{\bibfnamefont{M.~A.}
  \bibnamefont{Nielsen}}, \bibinfo{journal}{Phys. Rev. A}
  \textbf{\bibinfo{volume}{71}}, \bibinfo{eid}{062310} (\bibinfo{year}{2005}).

\bibitem[{\citenamefont{Rohde et~al.}(2005)\citenamefont{Rohde, Pryde, O'Brien,
  and Ralph}}]{ro-pra-72-032306}
\bibinfo{author}{\bibfnamefont{P.~P.} \bibnamefont{Rohde}},
  \bibinfo{author}{\bibfnamefont{G.~J.} \bibnamefont{Pryde}},
  \bibinfo{author}{\bibfnamefont{J.~L.} \bibnamefont{O'Brien}},
  \bibnamefont{and} \bibinfo{author}{\bibfnamefont{T.~C.} \bibnamefont{Ralph}},
  \bibinfo{journal}{Phys. Rev. A} \textbf{\bibinfo{volume}{72}},
  \bibinfo{eid}{032306} (pages~\bibinfo{numpages}{5}) (\bibinfo{year}{2005}).

\bibitem[{\citenamefont{Pryde et~al.}(2004)\citenamefont{Pryde, O'Brien, White,
  Bartlett, and Ralph}}]{pr-prl-92-190402}
\bibinfo{author}{\bibfnamefont{G.~J.} \bibnamefont{Pryde}},
  \bibinfo{author}{\bibfnamefont{J.~L.} \bibnamefont{O'Brien}},
  \bibinfo{author}{\bibfnamefont{A.~G.} \bibnamefont{White}},
  \bibinfo{author}{\bibfnamefont{S.~D.} \bibnamefont{Bartlett}},
  \bibnamefont{and} \bibinfo{author}{\bibfnamefont{T.~C.} \bibnamefont{Ralph}},
  \bibinfo{journal}{Phys. Rev. Lett.} \textbf{\bibinfo{volume}{92}},
  \bibinfo{pages}{190402} (\bibinfo{year}{2004}).

\bibitem[{\citenamefont{Ralph et~al.}(2006)\citenamefont{Ralph, Bartlett,
  O'Brien, Pryde, and Wiseman}}]{ra-pra-73-012113}
\bibinfo{author}{\bibfnamefont{T.~C.} \bibnamefont{Ralph}},
  \bibinfo{author}{\bibfnamefont{S.~D.} \bibnamefont{Bartlett}},
  \bibinfo{author}{\bibfnamefont{J.~L.} \bibnamefont{O'Brien}},
  \bibinfo{author}{\bibfnamefont{G.~J.} \bibnamefont{Pryde}}, \bibnamefont{and}
  \bibinfo{author}{\bibfnamefont{H.~M.} \bibnamefont{Wiseman}},
  \bibinfo{journal}{Phys. Rev. A} \textbf{\bibinfo{volume}{73}},
  \bibinfo{eid}{012113} (\bibinfo{year}{2006}).

\bibitem[{\citenamefont{Pittman and Franson}(2003)}]{pi-prl-90-240401}
\bibinfo{author}{\bibfnamefont{T.~B.} \bibnamefont{Pittman}} \bibnamefont{and}
  \bibinfo{author}{\bibfnamefont{J.~D.} \bibnamefont{Franson}},
  \bibinfo{journal}{Phys. Rev. Lett.} \textbf{\bibinfo{volume}{90}},
  \bibinfo{eid}{240401} (\bibinfo{year}{2003}).

\end{thebibliography}
\end{document}